\documentstyle[12pt,prl,aps]{revtex}
\input epsf

\draft
\begin{document}
\title{ 
Numerical Solutions of the von Karman Equations for a Thin Plate}
\author{Pedro Patr\'{\i}cio da Silva and Werner Krauth}
\address{
CNRS-Laboratoire de Physique Statistique de l'ENS\\
24, rue Lhomond; F-75231 Paris Cedex 05; France\\
e-mail: patricio@physique.ens.fr; krauth@physique.ens.fr\\
}
\date{September, 1996}
\maketitle
\begin{abstract}
In this paper, we present an algorithm for the solution  of the von Karman
equations of elasticity theory and related problems. Our method
of successive reconditioning is able to avoid convergence problems at
any ratio of the nonlinear streching and the pure bending energies.  
We illustrate the 
power of the method by numerical calculations of pinched
or compressed plates subject to fixed boundaries.
\end{abstract}
\pacs{PACS numbers: 02.60.Pn 02.70.Dn 46.30.Cn }
\newpage
The Navier-Stokes equation is not the 
only nonlinear partial differential equation in classical mechanics
which has so far eluded a systematic analytical or numerical solution.

Notoriously difficult partial differential equations appear also in the 
theory of elasticity; 
an example is given by the von Karman equations, which describe
the elastic energies  governing 
the deformation of a thin plate \cite{Landau}. 
The von Karman equations are simpler than 
the Navier-Stokes equation in 
that they are local, though nonlinear. A variational
theorem applies, which expresses the fact that the plate searches to minimize
the total elastic energy.

For a plate (described by the parametrization $\mathbf{r}$$(s,t)=(x,y,z)$)
of thickness $h$, the von Karman energy is given by \cite{Landau}
\begin{equation}
\int\int\{\frac{E_{el} h^{3}}{24(1-\sigma^{2})}\{(\Delta z)^{2}-2(1-\sigma)
[z,z]\}+\frac{h}{2}u_{ij}\sigma_{ij}\}dsdt
\end{equation}
where
\[[z,z]=\frac{\partial^{2}z}{\partial s^{2}}
\frac{\partial^{2}z}{\partial t^{2}}-(
\frac{\partial^{2}z}{\partial s\partial t})^{2}\]
$u_{ij}$ is the strain tensor with nonlinear terms in the deformations 
$u_{1}=x(s,t)-s$, $u_{2}=y(s,t)-t$, $z=z(s,t)$:
\[u_{ij}=\frac{1}{2}(\frac{\partial u_{i}}{\partial s_{j}}+
\frac{\partial u_{j}}{\partial s_{i}})+
\frac{1}{2}\frac{\partial z}{\partial s_{i}}
\frac{\partial z}{\partial s_{j}}\]
$\sigma_{ij}$ is the stress tensor, linearly proportional to $u_{ij}$. 
$E_{el}$ and $\sigma$ are the Young modulus and the Poisson ratio,
respectively.

The appropriate framework for a precise numerical solution of this 
elastical problem is the finite element method \cite{Schwarz}, 
in which the plate (described by intrinsic variables $s$ and $t$)
is cut up into triangles, the corners of which come to lie at specified
``nodes''. 
Inside each triangle, any of the three functions is interpolated 
by a high-order polynomial, such that it is continuously 
differentiable as one passes from one triangle to the next one.
This ensures a well-defined energy.
Classic work on the finite element method has established the 
existence of a ``magic'' polynomial of order $5$:
\begin{equation}
x(s,t)= a_1 + a_2 s + a_3 t + a_4 s\:t + \ldots + a_{21} t^5
\label{fifth}
\end{equation}
(and similarly for $y$ and $z$).
The  $21$ parameters in eq.~(\ref{fifth})
are fixed by prescribing at the three corners
the function value $x$, the first and second derivatives 
$\partial x / \partial s$, 
$\partial x / \partial t$, 
$\partial^2 x / \partial s^2$, $\partial^2 x / \partial s\partial t$, 
$\partial^2 x / \partial t^2$ as well as the normal derivatives $
\partial x / \partial n$ on the midpoints of the triangle's sides.
We denote these variables of different
origin by a symbol $\xi = (\xi_1, \ldots \xi_N)$, with $N$ the total number
of variables.  
The polynomial eq.~(\ref{fifth}) possesses an important symmetry property 
with respect to spatial rotations \cite{Schwarz}.
It uses all the 21 polynomial parameters to satisfy, in an optimal 
choice, the continuity of the first derivatives across the sides of the
triangle.	
Notice that  all the second derivatives of the 
functions are imposed at the nodes.
By construction, the functions $x(s,t)$ 
{\em etc}, are thus twice continuously differentiable at these points.

Given the nodes, and the variables $\xi$, it is straightforward to 
compute the interpolating 
polynomials (in the simplest fashion by inverting a 
$21 \times 21$ matrix), and to compute the {\em local} energy 
$E(s(\xi),t(\xi))$ 
as well as the {\em total} energy ${\cal E}(\xi)= 
\int E(s(\xi),t(\xi)) ds\;dt$ exactly
\cite{gaussfoot}.

As explained before, the finite element functions are 
valid variational test functions. 
It is therefore appropriate to search for the set of variables 
$\xi$ minimizing
the total energy. This multidimensional minimization problem (solve for
$ min_\xi {\cal E}(\xi)$) is 
at the heart of our present concern, and our main result will consist
in an algorithm which, for the first time,  makes possible  
a direct attack at the von Karman equations in the presence of strong
deformations, and at any value of the thickness \cite{Litt}.  
Any naive attempt to solve it is doomed to fail because
of the large number of variables at hand. In addition, we have to solve
the variational problem  to great precision (essentially achieve $|\nabla
{\cal E}| = 0$), 
since we are not really interested in the 
numerical value of the total energy, but in the geometrical aspects
of the solution, which are much slower to converge than $ {\cal E}$.

A few algorithms are specifically geared at the solution of large
minimization problems (for continuously differentiable functions).
They all attempt a local fit of the function by a parabola
\begin{equation}
 {\cal E}(\xi) \sim c + b\; \xi + \frac{1}{2} \xi H \xi
\label{parabola}
\end{equation}
where $H$ is the Hessian matrix of second derivatives. Based on the
knowledge of the function and of its numerical gradient $b$, strategies differ
on how to economize on the computation of $H$ \cite{Recipes}.

We have initially been extremely frustrated with the performance of these
algorithms for large $N$, especially in the strongly nonlinear 
regime of small $h$. 
We illustrate the  difficulties on a test example with $N=272$, 
a compressed
half cylinder ({\em cf} fig.~2) of thickness $h=0.01$, 
which will be further discussed later on. 
The upper curve in fig.~1 
shows the total energy ${\cal E}$ as a function of the iteration number, using 
one of the standard algorithms. For the given boundary 
condition, the simulation has proceeded for a few weeks on our 
work station without achieving reasonable convergence. In particular, 
the iteration never exhibits the quadratic convergence rate, which is
the hallmark of the fast minimization algorithms, and which allows in 
principle the solution of problems with hundreds or thousands of variables
\cite{Recipes}.

Before exposing our solution for the present nonlinear case ({\em i. e.} 
when $H$ depends on $\xi$), 
we shortly discuss the corresponding {\em linear} problem, in which 
the function $H$ is independent of $\xi$. The behavior of iterative 
minimization routines, such as the Hestenes-Stiefel conjugate 
gradient method, has been discussed in the literature, ({\em cf} \cite{Golub}
for a very clear discussion). The result is that the minimization 
algorithm performs well if the matrix $H$ is well conditioned or is
close to the unit matrix. Preconditioning algorithms have been 
devised, which transform the matrix $H$ into a similar matrix, 
close to the identity \cite{Golub}. 

The fact that the nonlinear minimization program in fig.~1 initially  
works very badly can thus only mean one of two things:  we either never 
enter the quadratic
region, where the energy $E(\xi)$ can be approximated
by eq.~(\ref{parabola}), or we have to do with 
badly conditioned matrices $H(\xi)$. It is an explicit computation 
of $H$ (which is not normally undertaken), that has convinced us on 
several examples that the approximation eq.~(\ref{parabola}) becomes acceptable
quite early in the simulation (it is to test this hypothesis that the
very long initial runs were undertaken). 
However, the matrix $H$ becomes 
very often extremely ill-conditioned especially if $h$ is small.
In the simulation followed in fig.~1, the 
eigenvalues  of $H$  span $6$ orders of magnitude at the point a)
(specified by variables $\xi_a$).

This has convinced us that a  {\em reconditioning} of the matrix 
is necessary. 
We have used the simplest reconditioning possible: every so often, say, at
point a), we explicitly compute $H$ ($ =H(\xi_a)$) and 
its eigenvalues $\alpha_i$  and eigenvectors $\eta_i$.
We then  choose {\em  new} variables $\tilde{\xi}=\eta_i/\sqrt{\alpha_i}$. 
At point a), the matrix $H$ thus ideally starts out as the unit matrix,  
and it then gets modified as we move away from $\xi_a$ \cite{renormfoot}. 

The expenditure of computing the matrix $H(\xi_a)$ and of reworking the 
complete calculation in terms of $\tilde{\xi}$ may seem enormous.
However, it is  immediately rewarded
by a strong decrease in the energy, and a concomitant  fast motion 
in the variable space. This can be seen on the middle  curve in fig.~1 
starting at point a). As we move away from $\xi_a$,  the condition of $H$ 
($ \equiv H(\xi)$) necessarily deteriorates again. In the example of fig.~1, 
starting from a well-conditioned matrix at point a), we reach 
at point b) an eigenvalue spectrum which again spans $5$ orders of
magnitude. In fact, most eigenvalues are of order $1$, but a few 
very small eigenvalues corrupt the condition of the matrix.

\begin{figure}[h]
\centerline{\epsfxsize=300pt\epsfbox{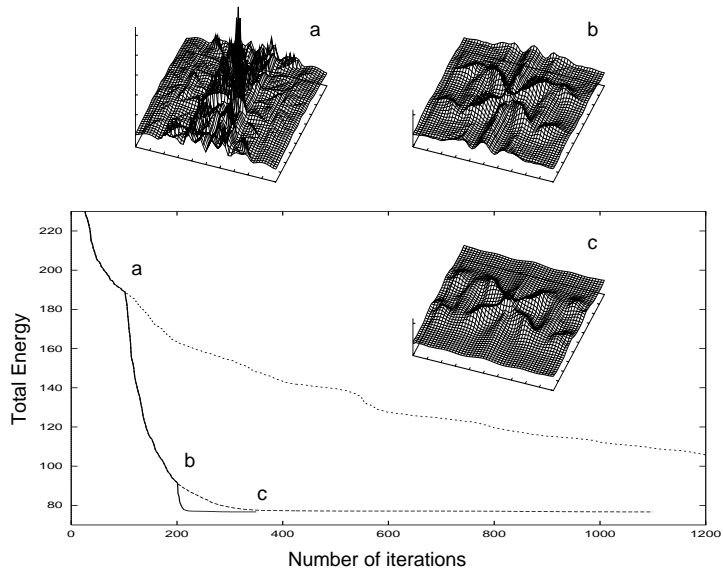}}
\caption
{Evolution of the minimization in a test case (same problem
as in fig.~2, $h=0.01$, $16$ nodes, $N = 272$ variables).
Upper curve:  
total energy ${\cal E}$ {\em vs} iteration using a standard
conjugate gradient algorithm. Reconditioning at a) and b) 
leads to the lower two curves. The insets show the 
local energy $ E(s,t)$ at  a), b) and c).
The final solution is given in fig.~2.}
\label{iteration}
\end{figure}

 If we continued our calculation, 
we would stay on the 
middle curve indicated in fig.~1, and undergo a gradual deterioration
of the convergence. If, on the other hand, we  recondition the matrix
a second time, at point b), we immediately approach the solution of the problem 
(and witness a quadratic convergence rate ({\em cf} \cite{Recipes}), 
which means that
we double the number of significant places of the solution {\em per} iteration).
A zero gradient of ${\cal E}$ (to machine precision) 
is reachable without problems. 

The final approach of the 
convergence is usually achieved after a few reconditionings, the 
precise number of which depends on the  physical nature of the 
problem. A good criterion is to try reconditioning if ${\cal E}$ 
no longer decreases even  though $|\nabla {\cal E}|$ is not yet
approaching $0$ (which would indicate convergence). It is thus 
evident that reconditioning is useless in the initial stage of the
minimization, in which even the standard algorithms decrease the energy
quite well.
In fig.~1, we also show the local energy distribution, at points
a), b) and c).

The physical problem discussed so far consists in an elastic plate, 
whose zero-energy configuration is the unit square. We impose a cylindrical
contour on two opposing sides. In addition, the plate is compressed
along these two sides, as indicated in fig.~2. In our numerical work, we
have to be concerned with the existence of local minima. In order to 
avoid the problem, we solve the minimization problem  repeatedly for 
increasing values of the compression.
For example, in the upper part of fig.~2, we first impose the cylindrical
shape, and then compute minima of the von Karman energy for increasing 
values of the lateral compression. For all the values of the plate 
thickness $h=0.1, \ldots, 0.001$ and of the compression, we have observed no 
numerical problems other than a gradual increase in the number of 
reconditionings required as $h$ decreases. 

\begin{figure}[h]
\centerline{\epsfxsize=300pt\epsfbox{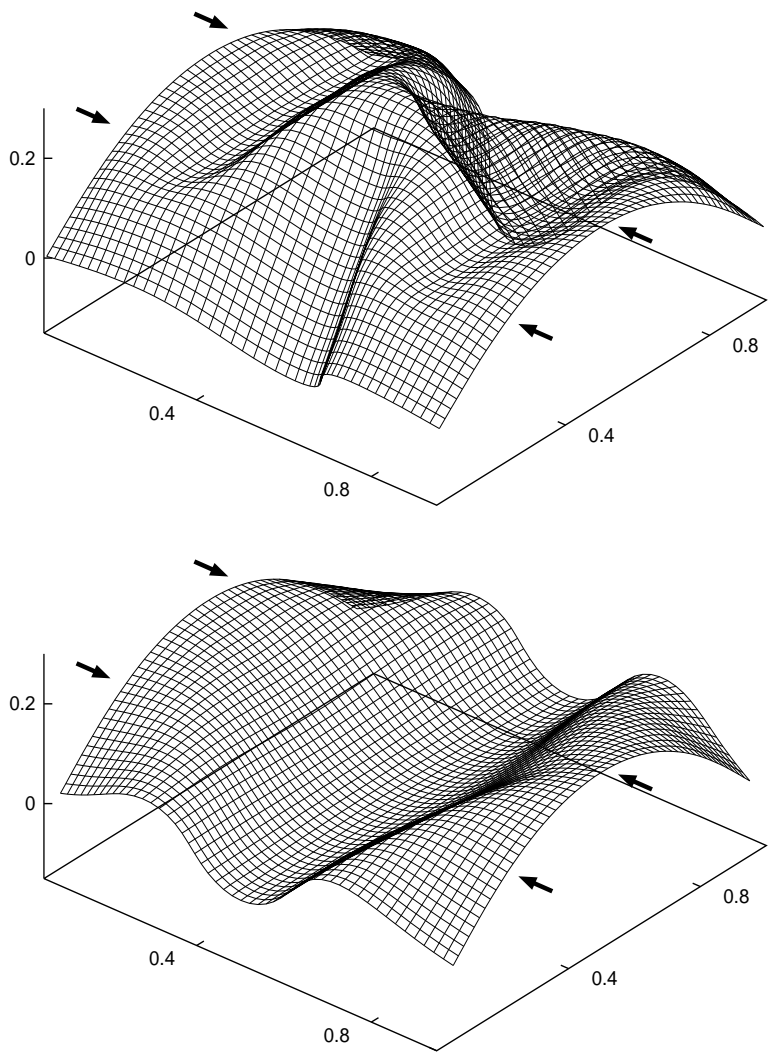}}
\caption
{Minimum energy solution for an elastic plate which was first bent, and then
compressed (upper) and first compressed and then bent (lower).}
\label{solution}
\end{figure}

In the lower part of the fig.~2, we have followed a different procedure, by
first compressing the flat plate, and then gradually imposing the 
cylindrical outer shape, in such a way that the final constraints are
exactly equivalent in the two parts of the figure. In both cases, we 
have undertaken extensive tests which have convinced us
that precise numerical solutions are obtained with of the order of
$16$ independent nodes, and a few hundred independent variables 
\cite{symmfoot}.

\begin{figure}[h]
\centerline{\epsfxsize=300pt\epsfbox{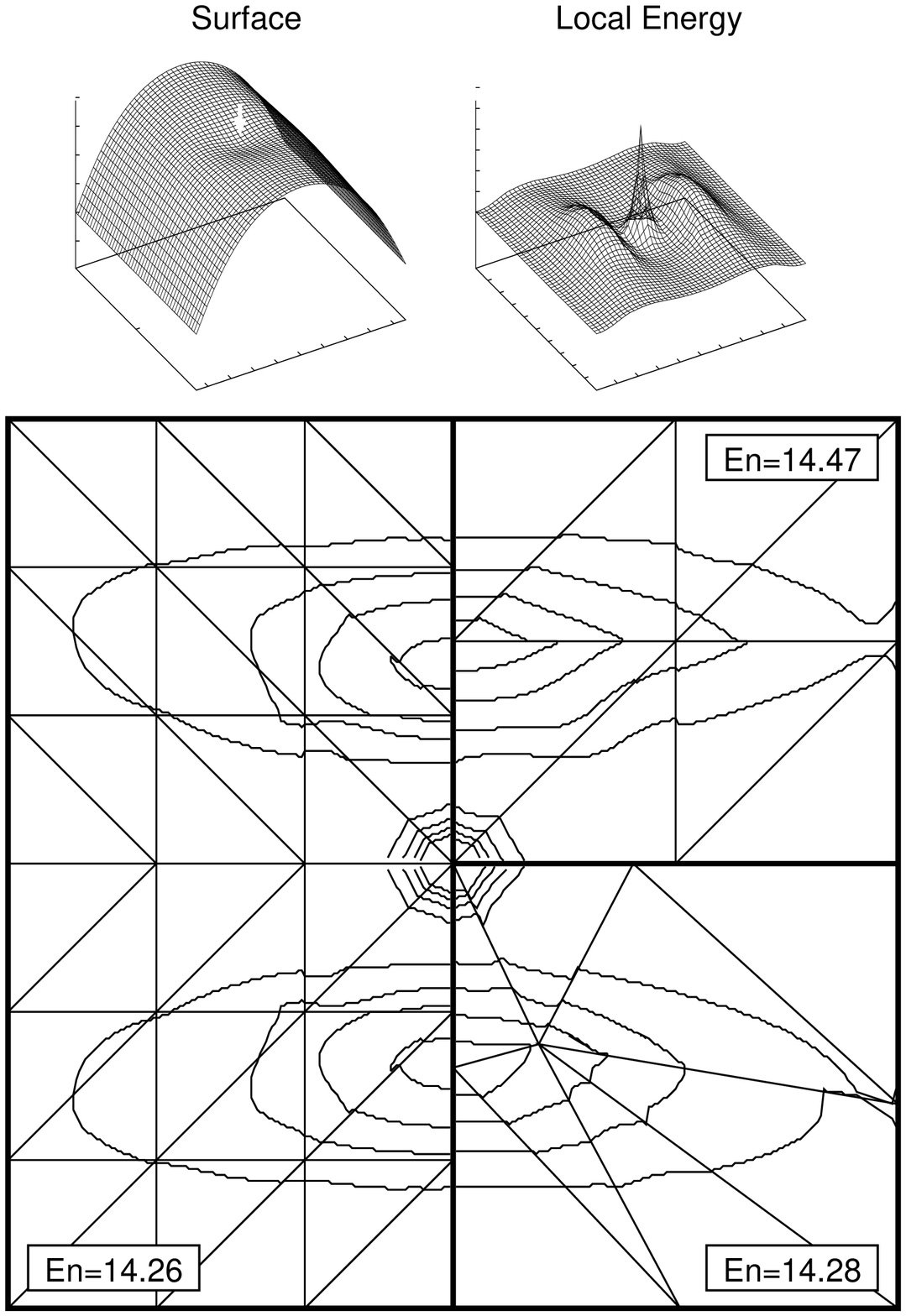}}
\caption
{Adaptive solution for the case of the pinched half-cylinder.
The main figure presents contour plots of $E(s,t)$ and the 
triangulation.\\
Left {\em two} quadrants: Nonadaptive calculation with 
$N=271$ variables, $16$ nodes.\\
Right upper quadrant: Nonadaptive, $N=126$, $9$ nodes\\ 
Right lower quadrant: Adaptive, $N=126$, $9$ nodes.\\
The adaptive calculation with $9$ nodes gives essentially 
the same variational energy as the non-adaptive calculation with 
$16$ nodes.
Iterative minimization was used,  as described in the text.}
\label{adaptive}
\end{figure}

Up to now, we have discussed the minimization problem only at fixed 
choice of the nodes. The most obvious way of increasing the
precision of a calculation consists in adding new ones, {\em e.g.}
by subdividing existing triangles. This strategy has been explored
in the literature \cite{Babuska}, but soon encounters its limits as
the number of variables increases very rapidly with the number of independent
nodes. 
However, the node positions may 
be included in the variational problem. Let us denote the node variables
by a symbol, $\eta$. A sharper variational function can be obtained not
only by increasing the dimension of $\eta$, but by solving 
the extended minimization problem 
\begin{equation}
   min_{\eta, \xi} {\cal E}(\xi(\eta))
\end{equation}
Just as the Gaussian integration (in which the function to be integrated is 
evaluated at optimal positions),
we can expect substantial gains in the quality of the 
variational functions. To show the usefulness of the approach, we present 
in fig.~3 the results of three separate numerical calculations on the 
problem of the half-cylinder which is pinched in the middle (we begin by
compressing the elastic plate into a half cylinder, and then 
impose a fixed position for the center of the plate, {\em cf.} the upper
left inset in fig.~3).
On the left, we show a 
contour plot of the energy density, 
and the position of the nodes for a (non-adaptive) numerical calculation 
with $N=271$ variables, and $16$ nodes. On the right upper side, 
we show the corresponding results
for a non-adaptive calculation with $N=126$ variables, $9$ nodes, 
and on the right lower side an {\em adaptive}
calculation, with the same number of nodes. There are two important 
features of the adaptive solution: the global energy ${\em E}$ is of the 
same quality as the much more expensive, non-adaptive solution with $16$ nodes.
Notice that the energy density resembles much more the one of the 
larger calculation, and that the nodes wander into regions of large
local energy.

In order to simplify the computation, we have, in fact not performed a 
full synchronous minimization over $\eta$ and $\xi$, which seems unnecessary.
In fact, it clearly appears that among the variables $\xi$, those which 
concern the function values, and the first derivatives converge 
much faster than the second derivatives (since the test functions  are
only once continuously differentiable). It is thus natural to suppose that
the change of the node positions will have the largest influence on second
derivatives. 
The minimization was done in an iterative form, in which the usual 
minimization ($ min_\xi {\cal E}(\xi)$ at fixed $\eta$)  is supposed to yield
reasonable values for the functions $x, y, z$, and the {\em first} derivatives.
At fixed function values and first derivatives, we then search for 
optimal values of $\eta$ and of the second derivatives, after which we
again  solve the usual problem, at the new $\eta$. This completely 
rigorous procedure (every function encountered is a true test function of
the von Karman energy) is then iterated several times. It quickly finds 
better nodes, which especially 
show smoother variations of the second derivatives of $x, y, z$ across 
the triangle boundaries.
 
In conclusion, we have discussed in this paper a very efficient method
to  solve numerically  the von Karman equations. In the examples studied,
our approach of successive 
reconditioning takes away all the  convergence difficulties (in the 
different regimes of nonlinearity, {\em i. e.} $h$). 
Further work will have to show whether the reconditioning can be 
obtained at 'reduced cost', without computing the full numerical 
Hessian.
We also discussed 
the idea of optimizing with respect to the node variables. This adaptive
choice of nodes was shown to be very useful. In the example, the 
nodes (and the sides of the triangle) migrate towards regions of 
very large local energy. We think it ultimately possible to 
take into account non-elastic terms in the von Karman equations, and 
thus produce a true numerical calculation of ``crumpled paper'' 
\cite{Benamar}. It should be evident that the method may be of general 
usefulness for non-linear minimization problems in high dimension.

Acknowledgment: We acknowledge helpful discussions with M. Ben Amar, 
who also got us interested in the present subject.

\noindent

\begin{references}
\bibitem{gaussfoot} The integrals, which are all polynomials in 
$s$ and $t$, can be computed exactly by Gaussian integration.
\bibitem{renormfoot} Due to numerical problems in determining $H$ and 
its eigensystem  for the very bad conditioning encountered, the new Hessian
$\tilde{H}$ is usually not very close to the unity matrix. 
The reconditioning can be iterated at the same point. 
The expense of doing this has to be weighted 
against the fact that $H$, of course, depends on $\xi$.
\bibitem{symmfoot} In both cases, we impose the symmetries $x\rightarrow -x$, 
$y\rightarrow -y$ of the solution.
\end{references}
\end{document}